\documentclass[11pt]{article}

\usepackage{amssymb,latexsym, amsmath}
\usepackage{graphicx}

\makeatletter
\@addtoreset{figure}{section}
\def\thefigure{\thesection.\@arabic\c@figure} \def\fps@figure{h, t}
\@addtoreset{table}{bsection}
\def\thetable{\thesection.\@arabic\c@table} \def\fps@table{h, t}
\@addtoreset{equation}{section}

\makeatother

\advance \topmargin by -\headheight
\advance \topmargin by -\headsep
\evensidemargin \oddsidemargin
\newtheorem{thm}{Theorem}[section]

\newfont{\tenbi}{cmbxti10}

\def\F{F}

\def\S{S}
\def\g{\mathfrak g}

\def\h{\mathfrak h}

\def\k{\mathfrak k}

\def\b{\mathfrak b}
\def\O{\mathfrak O}

\def\B{\mathbf B}
\def\G{\mathbf G}

\def\R{\mathbb R}

\begin{document}

\title{On billiard weak solutions of nonlinear PDE's and Toda flows
\footnote{PACS numbers 03.40.Gc, 11.10.Ef, 68.10.-m, AMS
Subject Classification 58F07, 70H99, 76B15}}
\author {Mark S. Alber
\thanks{Research partially supported by NSF grant DMS 9626672.} \\
Department of Mathematics\\
University of Notre Dame\\
Notre Dame, IN 46556 \\
{\footnotesize Mark.S.Alber.1@nd.edu}\\
\and
Roberto Camassa\\
Department of Mathematics\\
University of North Carolina\\
Chapel Hill, NC 27599\\
%{\small and}
%\\
%Center for Nonlinear Studies and
%Theoretical Division \\
%Los Alamos National Laboratory\\
%Los Alamos, NM 87545\\
{\footnotesize camassa@math.unc.edu}
\and
Michael Gekhtman\\
Department of Mathematics\\
University of Notre Dame\\
Notre Dame, IN 46556 \\
{\footnotesize Michael.Gekhtman.1@nd.edu}
}
\date{January 20, 1999}

\maketitle

\newpage

\begin{abstract}
A certain  class of partial differential  equations possesses
singular  solutions having discontinuous first derivatives
(``peakons").  The time evolution of peaks of such solutions is governed by a finite
dimensional completely integrable system. Explicit solutions of
this system are constructed by using algebraic-geometric method
which casts it as a flow on an appropriate Riemann surface and
reduces it to a classical Jacobi inversion problem. The algebraic 
structure of the finite dimensional flow is also examined in the
context of the Toda flow  hierarchy. Generalized peakon systems are
obtained for any simple Lie algebra and their complete
integrability is demonstrated.
\end{abstract}

%\tableofcontents

%%%%%%%%%%%%%%%%%%%%%%%%%%%%%%%%%%%%%%%%%%%%%%%%%%%%%%%%%%%%%%

\section{Introduction}
Camassa and Holm [1993] introduced and studied classes of
soliton-type weak solutions, for an integrable
nonlinear equation derived in the context of a shallow water
model.  In particular, they described the soliton dynamics in terms
of a system of Hamiltonian equations for the locations of the
``peaks" of the solution, the points at which its spatial
derivative changes sign. In other words, each peaked solution, 
or peakon,  can be associated with a mechanical system of moving
particles. New systems of this type were obtained in Calogero
[1995] and Calogero and Francoise [1996]. The $r$-matrix approach
was applied to the Lax pair formulation of an $n$-peakon system by
Ragnisco and Bruschi [1996], who also pointed out the connection 
of this system with the classical Toda lattice. A discrete version
of the Adler-Kostant-Symes factorization method was used by Suris
[1996] to study a discretization of the peakons lattice, realized
as a discrete integrable system on a certain Poisson submanifold of
${\rm gl}(n)$ equipped with $r$-matrix Poisson bracket.

In Alber {\it et al.} [1994, 1997, 1999] the existence of peakons
was linked to the presence of poles in the energy dependent
Schr\H{o}dinger operators associated with integrable evolution
equations. Namely, it was shown that the presence of a pole in the
potential is essential for a special limiting procedure which
allows for the formation of ``billiard' weak solutions. By using
the algebraic-geometric method, these billiard solutions are
related to finite dimensional integrable dynamical systems with
reflections. In this way,  profiles of billiard weak solutions are
associated with billiard motion inside quadrics with and without
the presence of a Hooke's potential. The points of impact on the 
quadrics
correspond to the  peaks of the profiles of weak solutions of 
the nonlinear PDE. Billiard solutions include new quasi-periodic and
soliton-like solutions, as well as peaked solitons with
compact support.
 This method can be used for
a number of equations including the shallow water equation, the Dym
type equation, as well as $N$-component systems with poles, 
together with all the equations in their respective hierarchies.

In Section 2 of this paper we derive equations for the motion of the peaks of
billiard solutions in the context of the algebraic-geometric
approach. We construct solutions of these equations by
connecting them to a flow on an appropriate Riemann surface
which leads to a classical Jacobi inversion problem. 

In Section 3 we show that orbits of the Toda flow in $sl(n)$ into
which the peakons lattice is mapped by means of the Lax matrix
introduced in Ragnisco and Bruschi [1996] admits a natural
generalization to the case of an arbitrary Lie algebra $\g$.
Restrictions of the Toda flow in $\g$ to these orbits can be viewed
as generalized peakons lattices. Using the methods developed in
Gekhtman and Shapiro [1999] we give an intrinsic description of
these orbits and prove complete integrability of the corresponding
flows. Unlike in the  $sl(n)$ case, Chevalley invariants of $\g$
are not sufficient to ensure integrability in the general situation,
and so we produce the requisite number of additional first
integrals. Lastly, we explain how to construct Darboux coordinates
for generalized peakon orbits and consider $sl(n)$ and $G_2$ as
examples.

\section{The Dynamics of the Peak Points}
\setcounter{equation}{0}
As shown in Alber {\it et al.} [1994, 1999], quasi-periodic solutions
of the shallow water equation
\begin{equation}
\label{SW}
U_t+3UU_x = U_{xxt}+2U_x U_{xx}+U U_{xxx}\,,
\end{equation}
on the infinite line
can be represented in the form
\begin{equation}\label{trace}
U(x,t)=\mu_1+\cdots+\mu_n - \mathfrak{m}
\end{equation}
where $\mu$'s are solutions
of the following systems of equations in $x$ and $t$,
\begin{equation} \label{mux}
\mu'_{i}=\frac{ \partial \mu_i}{\partial x}= \operatorname{Sign}
(\mu_i,m_{2i-1},m_{2i},s_i)
\frac{\sqrt{R(\mu_i)} } {\mu_i\prod_{j\ne i}^n(\mu_{i}-\mu_{j})}
\end{equation}
and
\begin{equation} \label{mut}
\dot{\mu}_{i}=\frac{ \partial\mu_i}{\partial t}
=\operatorname{Sign} (\mu_i,m_{2i-1},m_{2i},d_i)
\frac{(\mu_i-\Sigma)\sqrt{R(\mu_i)}}
{\mu_i\prod_{j\ne i}^n (\mu_{i}-\mu_{j}) },
\cr i=1,\dots,n, \quad \Sigma
=\mu_1+\cdots+\mu_n,
\end{equation}
with
\begin{equation} \label{poly1first}
R(\mu)=\mu\prod_{r=1}^{2n+1}(\mu-m_{r}) \, .
\end{equation}
Here the constants $m_r$ depend on the initial condition
for the equation (\ref{SW}), $\mathfrak{m}=\sum_r^{2n+1}m_r$,
and $\operatorname{Sign}(y,l,r,s)$
is a function on the Riemann surface,
$
\Gamma=\{w^2=R(\mu)\} \, ,
$
which switches from $-1$ to $1$ and back each time $y$ reaches
$l$ or $r$, endpoints of a cut on the Riemann surface.  These
ordinary differential equations (ODE's) for  the $\mu$-variables 
provide half of the equations of a finite dimensional
Hamiltonian system. Notice that periodic solutions form a subclass
of the quasi-periodic solutions. Also, soliton solutions can be obtained
from quasi-periodic solutions by shrinking pair wize the roots $m_r$ 
of the spectral polynomial $R(E)$.

Peaked quasi-periodic solutions of equation~(\ref{SW}) can be
constructed from solutions of (\ref{mux}) and (\ref{mut}) by using
the limiting process $m_1 \rightarrow 0$ (see Alber {\it et al.}
[1994,1999]) and the trace formula (\ref{trace}). In what follows ODE's
governing the time evolution of the peak locations are obtained and
a connection with the canonical variables of Camassa and Holm [1993] 
is described.

The $x$ and $t$ evolution  of the $\mu$ variables can also be connected to  a
constrained motion of a particle  on an $n$-dimensional hypersurface
imbedded  in
$\mathbb{R}^{n+1}$ which is parameterized by  the constants $m_r$, under the
action
of a harmonic force field (see Alber {\it et al.} [1999]).  The limit
$m_1\to 0$
corresponds to ``flattening" this surface along one direction which
results in
harmonically   forced motion in a region of $\mathbb{R}^{n}$ confined by an
$(n-1)$-dimensional boundary. The particle  collides with the boundary and moves
along segments of an
$n$-dimensional billiard.  The reflection on the boundary 
causes, after using the trace
formula~(\ref{trace}), the appearance of peaks 
in the corresponding  PDE solutions of
equation~(\ref{SW}).

On the level of    (\ref{mux}) and (\ref{mut}), the boundary of the
billiard  can be shown to be described by the following condition:
$\mu_1(x,t)=0$.
Reflection on the boundary is described by the sign switch from 1 to -1 (or
from -1 to
1) of first derivative of $\mu_1$ in (\ref{mux}) and (\ref{mut}).

\paragraph{The Periodic Case.}
The main steps in deriving
equations for the peak locations can be demonstrated by considering the
following basic
example. Let us assume
$n=2$, and
$m_1=0$.
Equation~$\mu_1(x,t)=0$ defines a function of time
$x=q(t)$. An equation for the time dependence of $q(t)$ can
be found by differentiating $(\mu_1(q(t),t)=0) $ with respect to time. We
have
\begin{equation}
\label{dmu1dt}
\frac{d}{dt}\Big(\mu_1(q(t),t) \Big) = 0 =
\frac{\partial \mu_1}{\partial x}(q(t),t) \dot{q}
+ \frac{\partial \mu_1}{\partial t}(q(t),t).
\end{equation}
The commuting  $x$- and $t$-flows of $\mu_1$ are described by
\begin{eqnarray}
&&\frac{\partial \mu_1}{\partial x}
=\operatorname{Sign}(\mu_1)
\frac{\sqrt{C_{4}(\mu_1)}}{\mu_1-\mu_2}\,,
\\
&&\frac{\partial \mu_1}{\partial t}
=\operatorname{Sign}(\mu_1)
\frac{(-\mu_2)\sqrt{C_{4}(\mu_1)}}{\mu_1-\mu_2},
\nonumber
\end{eqnarray}
$$C_{4}(\mu) = (\mu - m_2)(\mu - m_3)(\mu-m_4)(\mu-m_5).$$
Notice that the spatial derivative of $\mu_1$ is not defined at
$x=q(t)$, so that in the above formula
${\partial \mu_1}/{\partial x}$ evaluated at $(q(t),t)$ needs to be
found in the limit as $x\to q$ either from the left, or from the right.
The sign uncertainty connected with this choice factors out from the
final formula for $\dot{q}$. This combined with (\ref{dmu1dt})
and the billiard condition at the boundary $\mu_1(q(t),t)=0$ gives
\begin{equation}
\label{qm2dt}
 \dot{q}=\mu_2(q(t),t) \, .
\end{equation}
We also need an evolution equation for the quantity $\mu_2(q(t),t)$.
Notice that from the trace formula (\ref{trace}) and the 
expression for the boundary $\mu_1(q(t),t)=0$, this is the quantity
that determines the amplitude  of the  peak $U(q(t),t)$. The time
derivative of $\mu_2(q(t),t)$ is
\begin{equation}
\frac{d}{dt}\Big(\mu_2(q(t),t) \Big)
= \frac{\partial \mu_2}{\partial x}(q(t),t) \dot{q}
+ \frac{\partial \mu_2}{\partial t}(q(t),t) \, ,
\end{equation}
which used with the commuting $x$- and $t$-flows equations for $\mu_2$
yields, in analogy  with system (\ref{qm2dt}),
\begin{align}
\frac{dy}{dt}(t)
& = -\operatorname{Sign}(y)\frac{\sqrt{C_{4}(y)}}{y}\Big(\dot{q}+B_1(y)\Big)
\nonumber
\\
& =  -\operatorname{Sign}(y) \frac{\sqrt{C_{4}(y)}}{y} \dot{q} \, .
\label{dyt}
\end{align}
Here we have introduced the notation
$$
 y(t)=\mu_2(q(t),t)
$$
and used again the definition of $q(t)$ so that
\[
B_1(\mu_2(q(t))
\equiv B_1(y)=-\mu_1(q(t),t)=0 .
\]
 Combining this equation for
$\dot{y}$ with the evolution equation (\ref{qm2dt}) for $q$ gives
the following system for the variables $y$ and $q$,
\begin{eqnarray}
\label{yq}
&&\dot{y}=- \operatorname{Sign}(y) \sqrt{(y-m_2)(y-m_3)(y-m_4)(y-m_5)}
\nonumber \\
&&\dot{q}=y
\end{eqnarray}
Notice that the evolution for $y$ is decoupled from that of $q$.
The inversion problem corresponding to the equation for $y$ determines
a periodic function of $t$ with nonzero average. The evolution of $q$
is then given by the combination of a linear growth in time with slope
given by the average of $y$, plus periodic oscillations. We also
remark that for weak solutions of equation~(\ref{SW}) having isolated
discontinuities in the first derivative, the jump
condition yields
$$\frac{d q}{ d t} = U(q(t),t),$$ and using
the trace formula~(\ref{trace}) this can be seen to coincide with
the second equation in system~(\ref{yq}).

\paragraph{The Genus 2 Quasi-periodic Case.}
We describe a billiard genus 2 quasi-periodic solution. The spectral
polynomial in this
case is of the $5$th order and will be denoted as $C_5(\mu)$. The boundary
is again
described by
$\mu_1(q(t),t)=0$. We shall write
\[
 y_1 = \mu_2(q(t),t)
\quad  \mbox{and} \quad
y_2 = \mu_3(q(t),t) .
\]
 From the weak solution
approach it follows that
$\dot q = U(q_1) = y_1 + y_2$.
We differentiate $y_1$ and $y_2$ to obtain
\begin{equation}
\label{ys}
\left. \begin{array}{l}  {\displaystyle
\dot y_1 = \frac{d}{dt} \left [ \mu_2(q,t) \right ]
= \frac{\partial \mu_2}{\partial x} \dot q
+ \frac{\partial \mu_2}{\partial t}
= \frac{\operatorname{Sign}(\mu_2)
\sqrt{C_6(\mu_2)}}{\mu_2 (\mu_2 - \mu_3)}
(\dot q - \mu_3)
= \frac{\operatorname{Sign}(y_1)
\sqrt{C_6(y_1)}}{y_1 - y_2}}\\
\\ {\displaystyle
\dot y_2 = \frac{d}{dt} \left [ \mu_3(q,t) \right ]
= \frac{\partial \mu_3}{\partial x} \dot q
+ \frac{\partial \mu_3}{\partial t}
= \frac{\operatorname{Sign}(\mu_3)
\sqrt{C_6(\mu_3)}}{\mu_3 (\mu_3 - \mu_2)}
(\dot q - \mu_2)
= \frac{\operatorname{Sign}(y_2)
\sqrt{C_6(y_2)}}{y_2 - y_1}.}\\
\end{array} \right \}
\end{equation}
This is a well-defined system associated with a genus $2$ Riemann
surface. The corresponding problem of inversion involves only
holomorphic differentials and therefore the quantities $y_1$,$y_2$ and
$q(t)$ can be expressed in terms of standard $\theta$-functions on the
Riemann surface.

%%%%%%%%%%%%%%%%%%%%%%%%%%%%%%%%%%%%%%%%%%%%%%%%%%%%%%%%%%%%%%
\paragraph{Peakon solutions.}
We now specialize the above formalism to the limiting case of soliton
solutions of equation~(\ref{SW}) on the real line, in which
each pair $m_{2i},m_{2i+1}$ is taken to limit to the
$a_i$, $i=1, \dots, n$. 
For just one $\mu$-variable,
using the trace formula $U = \mu -a$ at the billiard boundary
results in $U(q(t),t) = -a$.
The single peakon solution determined by this procedure is therefore
\begin{equation}
\label{peakon}
U(x,t) = -a e^{-|x+a t|}\,,
\end{equation}
which is a traveling wave soliton-type solution.
Camassa and Holm [1993] found
\begin{equation} \label{peakonr}
U(x,t) = p(t) e^{-|x-q(t)|}\,,
\end{equation}
with an additional link between $p(t)$ and $q(t)$ (Hamiltonian
structure) which in the $1$-peakon case results in $p(t)$ being a
constant and $q(t)$ being a linear function.

%%%%%%%%%%%%%%%%%%%%%%%%%%%%%%%%%%%%%%%%%%%%%%%%%%%%%%%%%%%%%%

\paragraph{2-peakon solution.} In this case, one obtains the following
system that describes $2$-peakon profiles,
\begin{equation} \label{sw2}
\left. \begin{array}{l} {\displaystyle
\frac{\partial \mu_1}{\partial x} =
\operatorname{Sign}(\mu_1) \frac{(\mu_1 - a_1)(\mu_1 - a_2)}{(\mu_1 - \mu_2)}}
\\
\\ {\displaystyle
\frac{\partial \mu_2}{\partial x} =
\operatorname{Sign}(\mu_2) \frac{(\mu_2 - a_1)(\mu_2 - a_2)}{(\mu_2 - \mu_1)}}
\end{array} \right \}
\end{equation}
where $\mu_1$ and $\mu_2$ are evaluated between $a_1$ and $0$ and
$a_2$ and $0$, respectively. The corresponding time-flow is
\begin{equation} \label{sw2t}
\left. \begin{array}{l} {\displaystyle
\frac{\partial \mu_1}{\partial t} =
\operatorname{Sign}(\mu_1)B_1(\mu_1)\frac{(\mu_1 - a_1)(\mu_1 -
a_2)}{(\mu_1 - \mu_2)}}
\\
\\ {\displaystyle
\frac{\partial \mu_2}{\partial t} =
\operatorname{Sign}(\mu_2) B_1(\mu_2)\frac{(\mu_2 - a_1)(\mu_2 -
a_2)}{(\mu_2 - \mu_1)}}
\end{array} \right \}
\end{equation}
and the first order polynomial $B_1$
for equation~(\ref{SW}) gives
$$
B_1(\mu_1)=-\mu_2+a_1+a_2, \qquad B_1(\mu_2)=-\mu_1+a_1+a_2\, .
$$
Define the boundary by introducing functions $q_1(t)$ and $q_2(t)$
such that
\begin{equation} \label{q12}
\mu_1 (q_1(t),t) = 0,\;\;\mu_2 (q_2(t),t) = 0\,.
\end{equation}
Define functions
\begin{equation} \label{y12first}
y_1 = \mu_2 (q_1(t),t),\;\;y_2 = \mu_1 (q_2(t),t).
\end{equation}
Differentiating $\mu_1 (q_1(t),t)$ and $\mu_2 (q_2(t),t)$ in~(\ref{q12})
results in
\begin{equation}\label{mu12dot}
\left. \begin{array}{l} {\displaystyle
\frac{d \mu_1}{d t} = 0 =\frac{\partial \mu_1}{\partial x}
\left(
\frac{d q_1}{d t} +B_1(\mu_1) \right)= \frac{\partial \mu_1}{\partial x}
\left(\frac{d q_1}{d t}-y_1+a_1+a_2 \right)  }
\\
\\
{\displaystyle
\frac{d \mu_2}{d t} = 0 =\frac{\partial \mu_2}{\partial x}
\left(
\frac{d q_2}{d t} +B_1(\mu_2) \right)= \frac{\partial \mu_2}{\partial x}
\left(\frac{d q_2}{d t}-y_2+a_1+a_2 \right)  }\end{array} \right \}
\end{equation}
which coincides with the jump conditions for weak solutions
\begin{equation}
\label{xeqn}
\left. \begin{array}{l} {\displaystyle
\frac{d q_1}{d t} = U(q_1)
= \left ( \mu_1 + \mu_2 - a_1 - a_2 \right ) \big|_{q_1}
= \mu_2 (q_1) - a_1 - a_2} = y_1 - a_1 - a_2
\\
\\ {\displaystyle
\frac{d q_2}{d t} = U(q_2)
= \left ( \mu_1 + \mu_2 - a_1 - a_2 \right ) \big|_{q_2}
= \mu_1 (q_2) - a_1 - a_2 = y_2 - a_1 - a_2} \, .
\end{array} \right \}
\end{equation}
Finally, differentiate $y_1$ and $y_2$ to find
$$
\left. \begin{array}{l} {\displaystyle
\frac{d y_1}{d t} = \frac{\partial \mu_2}{\partial x}
\frac{d q_1}{d t} + \frac{\partial \mu_2}{\partial t}
=\frac{\partial \mu_2}{\partial x}
\left(\frac{d q_1}{d t}+B_1(\mu_2)\right)}
\\
\\ {\displaystyle
\frac{d y_2}{d t} = \frac{\partial \mu_1}{\partial x}
\frac{d q_2}{d t} + \frac{\partial \mu_1}{\partial t}=
\frac{\partial \mu_1}{\partial x}
\left(\frac{d q_2}{d t}+B_1(\mu_1)\right)}
\end{array} \right \}
$$
and so
\begin{equation} \label{y12}
\left. \begin{array}{l} {\displaystyle
\frac{d y_1}{d t}
=\operatorname{Sign}(y_1)(y_1 - a_1)(y_1 - a_2) }
\\
\\ {\displaystyle
\frac{d y_2}{d t}
=\operatorname{Sign}(y_2) (y_2 - a_1)(y_2 - a_2)}
\, .
\end{array} \right \}
\end{equation}
Thus, the equations of evolution for $y_1$ and $y_2$ decouple from
those of $q_1$ and $q_2$. The decoupled
equations~(\ref{y12})  can be solved first
and $q_1$, $q_2$ subsequently determined from~(\ref{xeqn})
by quadratures.

It is interesting to examine the connection of the  set of
variables $q_i,y_i$, $i=1,2$, with the $q_i,p_i$, $i=1,2$,
introduced by Camassa and Holm [1993]. By definition, the $q$'s are
the same in both sets (the $(q,y)$ system does have a
canonical Hamiltonian form).
As to the $y$'s, notice that
\begin{equation} \label{ypq}
\left. \begin{array}{l} {\displaystyle
U(q_1)=y_1 - a_1 - a_2=p_1+p_2 e^{-|q_1-q_2|}}
\\
\\
{\displaystyle
U(q_2)=y_2 - a_1 - a_2=p_2+p_1 e^{-|q_1-q_2|}}
\end{array} \right \}
\end{equation}
which provides a definition of the transformation of the
$y$ variables   in terms of $p$'s and $q$'s. This together with~(\ref{xeqn})  yields
the first set of equations for the Hamiltonian system
derived by Camassa and Holm [1993],
\begin{equation} \label{qqdot}
\left. \begin{array}{l}
{\displaystyle
\frac{d q_1}{d t}=p_1+p_2 e^{-|q_1-q_2|}}
\\
\\
{\displaystyle
\frac{d q_2}{d t}=p_2+p_1 e^{-|q_1-q_2|}}
\, .
\end{array} \right \}
\end{equation}
The constants of motion
$a_1$ and $a_2$ can be expressed in terms of $p$'s and $q$'s
via the first integrals
$$
P_{12}=p_1+p_2=-(a_1+a_2),
$$
and
$$
H_{12}=\frac{1}{2}(p_1^2+p_2^2)+p_1 p_2
e^{-|q_1-q_2|}=\frac{1}{2}(a_1^2+a_2^2)\, ,
$$
which are the total momentum and Hamiltonian
for the $(q,p)$-flow, respectively.
Using these expressions, the constants of
motion $a_1$ and $a_2$ can be eliminated from~(\ref{ypq})
and the explicit variable transformation of variables for
$y_1$, $y_2$ in terms of the $(q,p)$ system is
\begin{equation} \label{ppqqy}
\left. \begin{array}{l}
{\displaystyle y_1=p_2 ( e^{-|q_1-q_2|}-1) }
\\
{\displaystyle y_2=p_1 ( e^{-|q_1-q_2|}-1) }
\, .
\end{array} \right \}
\end{equation}
Differentiating these equations with respect to time,
using~(\ref{y12}),  (\ref{xeqn}), and the
transformation~(\ref{ppqqy}) itself, results in a
system of equations for $\dot{p}_1$ and $\dot{p}_2$,
which can be solved to yield
$$
\left. \begin{array}{l}
{\displaystyle
\frac{d p_1}{d t}=\operatorname{sgn}(q_2-q_1) p_1p_2 e^{-|q_1-q_2|}}
\\
\\
{\displaystyle
\frac{d p_2}{d t}=\operatorname{sgn}(q_1-q_2)p_2p_1 e^{-|q_1-q_2|}}
\, .
\end{array} \right \}
$$
Equations~(\ref{qqdot}) and~(\ref{ppqqy})  are precisely
those that follow from $H_{12}$ with a canonical
Hamiltonian structure in the $(p,q)$ variables.

The case of
three or more  derivative-shock singularities $x_i,y_i$, $i=1,2, \dots, N
\geq 3$ proceeds in complete analogy with the case $N=2$ above.
Once again, the
$q$-flows decouple from those of the $y$'s, however the
equations in the system governing the $y$-flow are now coupled
and it is not immediately obvious that this system is integrable.
A closer inspection
however reveals that the $y$-flow shares the same structure
as that of the $\mu$-variables flow and is therefore integrable
by a similar argument.

\section{Generalized Peakons Lattices}
In this section we discuss the connection between the peakons lattice and
the Toda flows. The fact that the peakons lattice can be realized
as a special case of the Toda flows in $gl(n)$ was discovered
in Ragnisco and Bruschi [1996] and later used by Suris [1996]
to introduce a discrete time peakons lattice. Let us review
these results, concentrating, for simplicity, on a particular
case of the  peakons lattice with a Hamiltonian
\begin{equation}
H(p,q)= \frac{1}{2} \sum_{i=1}^{n}p_i^2 + \sum_{1\le i<j\le
n}p_ip_j\exp(q_j-q_i)\ ,
\label{pl}
\end{equation}
that corresponds to the case when particles $q_i$ are ordered as follows:
$q_1>q_2>\ldots>q_n$. It turns out that (\ref{pl}) possesses a Lax
representation with the Lax matrix
\begin{equation}
L=L(p,q)= \sum_{i=1}^{n}p_i E_{ii} + \sum_{1\le i<j\le n}\sqrt{p_ip_j}
\exp\left ( \frac{1}{2} (q_j-q_i) \right ) (E_{ij} + E_{ji}),
\label{laxpl}
\end{equation}
and an auxiliary matrix equal to the half of the skew-symmetric part of $L$,
the peakons lattice thus being a restriction of the Toda flow to
the set of matrices of the form (\ref{laxpl}).
It was shown that this set forms a Poisson
submanifold w.r.t. the associated $r$-matrix bracket on $gl(n)$. Complete
integrability of  (\ref{pl}) is provided by the Chevalley invariants of
$gl(n)$, i.e.
spectral invariants of $L$.

In what follows we give a uniform Lie-algebraic procedure which
allows us to construct peakon-type orbits in any simple Lie algebra.
On each of these orbits the Toda flow is shown to be  a completely integrable
system. Notice that for proving this we provide additional first integrals
which supplement the Chevalley invariants of the algebra. We also indicate a
convenient
way of constructing Darboux coordinates for orbits under  consideration.

Let us first recall the Hamiltonian formalism for the generalized (symmetric)
Toda flows (cf. Kostant [1979], Goodman and Wallach [1985],
Reyman and Semenov-Tian-Shansky [1994]).
Let $\g$ be the normal real form of a simple Lie algebra of rank $r$, $\G$  a
corresponding Lie group and  $\h$ a  Cartan subalgebra of $\g$.
We denote by $\Phi$ the root system of $\g$, and  by
$\Phi^+$ (resp. $\Phi^-$) the set of all positive (resp. negative)
roots. We also fix a Chevalley basis
$\{ e_\alpha,\ \alpha\in\Phi; \ h_i,\ i=1,\dots, r\}$ in $\g$ .
All properties of the root systems and Chevalley bases that we
need can be found in Humphreys [1980] and Onishchik and Vinberg [1990].

Consider a direct sum decomposition $\g= \k + \b_+$, where $\b_+$ is the
upper Borel subalgebra and $\k$ is  the maximal compact subalgebra of $\g$.
The dual space $\b^*_+$ of $\b_+$ can be identified
with the space $\S$ of ``symmetric''
elements of $\g$ that have a form
\begin{equation}
L=\sum_{j=1}^rb_jh_j+\sum_{\alpha\in\Phi}a_\alpha(e_{\alpha}+e_{-\alpha})\, .
\label{symm}
\end{equation}
$\S$ is an annihilator of $\k$ w.r.t the Killing form $\langle\ , \ \rangle$.

The pull-back of the Lie-Poisson bracket on $\b^*_+$ equips $\S$ with the
Poisson bracket
\begin{equation}
\{f_1, f_2\}_{\S} (L)=
\frac{1}{2}\langle L , [\pi_+ \nabla f_1 (L), \pi_+ \nabla f_2 (L)] \rangle\ ,
\label{Sbrack}
\end{equation}
where gradients are defined w.r.t. the Killing form and
$\pi_+$ is a projection on  $\b_+$ parallel to $\k$.
Equations of motion of the generalized Toda flow on $\S$ are generated by a
Hamiltonian $H(L)={1\over 2}\langle L,L\rangle$ and have the Lax form
\begin{equation}
\dot L=[L,\frac{1}{2}\pi_+(L)]=[\frac{1}{2}\pi_{-}(L), L]\
\label{Lax}
\end{equation}
where $ \pi_{-} = Id - \pi_+$ . The bracket (\ref{Sbrack}) is
a restriction to $\S$ of the so called $r$-matrix bracket on $\g$:
\begin{equation}
\{f_1, f_2\}_{r} (X)=
\frac{1}{2}\langle X , [\pi_+ \nabla f_1 (X), \pi_+ \nabla f_2 (X)] -
[\pi_- \nabla f_1 (X), \pi_- \nabla f_2 (X)] \rangle
\ .
\label{Rbrack}
\end{equation}

Symplectic leaves of the Poisson manifold $( S, \{ , \}_{\S})$
coincide with the orbits of the coadjoint action of the upper Borel
subgroup $\B_+$ of $\G$:
\begin{equation}
\mbox{Ad}_b^*(L)= \pi_S \mbox{Ad}_{b^{-1}}(L)\, ,
\label{coad}
\end{equation}
where $\pi_S$ is a projection on  $S$ parallel to $\b_+$.
On every indecomposable symplectic leaf, restrictions of the Chevalley
invariants $I_1=H, I_2,\ldots, I_r$ of $\g$ form a family
of independent Poisson commuting integrals for the Toda flow. This family is
maximal, however, only for a few distinguished orbits, that were classified
in Goodman and Wallach [1984] and Perelomov and Kamalin [1985]

To describe a particular type of coadjoint orbits, which in the $sl(n)$
case will be
shown to be associated with the peakons lattice, we need the following
definitions
introduced in Gekhtman and Shapiro [1999].
Let $m$ be the maximal positive root and
\begin{equation}
h_m= [e_m, e_{-m}] \ .
\label{hm}
\end{equation}
Define
\begin{eqnarray}
\nonumber
&\g'=\mbox{Ker}_{\mbox{ad}_{e_m}}\cap \mbox{Ker}_{\mbox{ad}_{e_{-m}}}\\
&\F= \mbox{Span} \{ e_\alpha : \alpha \in \Phi^- \ \mbox{and}\ (
m,\alpha )
\ne 0\}\\
\nonumber
&\tilde \F = \mbox{Span} \{ \F , h_m\}\\
\nonumber
&V =\F - \R \langle e_{- m} \rangle \ .
\end{eqnarray}
Then $\g'$ is a semisimple subalgebra of $\g$ and $\F$ is a  Heisenberg
subalgebra of $\g$, i. e. $V$ is spanned by root vectors $e_{\alpha_i},
e_{-m-\alpha_i},\ i=1,\ldots N$, such that
\begin{eqnarray}
\nonumber
&[e_{\alpha_i}, e_{\alpha_j}]=[e_{-m-\alpha_i}, e_{-m-\alpha_j}]=
[e_{-m},e_{\alpha_i}]=[e_{-m},e_{-m-\alpha_i}]= 0,\\
&[e_{\alpha_i},
e_{-m-\alpha_j}]=c_i \delta_{i}^{j} e_{- m},\ ,
\label{rel1}
\end{eqnarray}
where $c_i$ are some positive  constants. Note also that
\begin{equation}
[h_m, e_{\alpha_i}]= - e_{\alpha_i},  \ [h_m, e_{-m-\alpha_i}]= -
e_{-m-\alpha_i} ,\
[h_m,e_{\pm m}]=\pm 2 e_{\pm m} \ ,
\label{rel2}
\end{equation}
Therefore (\ref{Sbrack}) and (\ref{rel2}) yield the following Poisson brackets
\begin{equation}
\{ y_0,x_0 \}_\S\ = x_0,\
\{ y_0, x_i \}_\S\ =\frac{1}{2}x_i,\ \{ y_0, y_i \}_\S\ =\frac{1}{2}y_i,\
\{ y_i, x_i\}_\S\ = \frac{1}{2}x_0,\ i=1,\ldots N,
\label{rel3}
\end{equation}
for the linear functions
\begin{equation}
x_i= \langle L, e_{-\alpha_i}\rangle,\
y_i= \frac{1}{c_i}  \langle L, e_{m+\alpha_i}\rangle,\
x_0=\langle L, e_{m}\rangle,\
y_0=\langle L, h_m \rangle\ .
\label{linear}
\end{equation}
All other brackets between functions (\ref{linear}) are zero.

Let $h_0\in\h$ be orthogonal to $h_m$ w.r.t the Killing form and let
$\O_{m, h_0}$ be the coadjoint orbit of $\B_+$ through $(e_{-m} + h_0 + e_m)$.

\begin{thm} $\O_{m, h_0}$ can be parameterized by elements of $\tilde \F$ :

\begin{equation}
\O_{m, h_0} = \left  \{ L= \pi_\S \left( \zeta - \frac{1}{2 x(\zeta)}
[\mbox{ad}_{e_m} v(\zeta), v(\zeta)] \right)\ : \zeta\in \tilde \F_-,
x(\zeta) >0
\right \}\ ,
\label{orbit}
\end{equation}
where element $\zeta\in \tilde \F_-$ is decomposed as
$$\zeta= x(\zeta) e_{-m} + y(\zeta) h_m + v(\zeta)\ ,
$$
and $v(\zeta)\in V.$

The Toda flow (\ref{Lax}) is completely integrable on $\O_{m, h_0}$ with
a maximal family of functions in involution given
by the coefficients of polynomials
\begin{equation}
I_j \left(L + \frac{\lambda}{x(\zeta)}e_m \right)=\sum_k
I_{jk}(L)\lambda^k,\ j=1,\ldots r,
\label{ij}
\end{equation}
where $I_j$ are the Chevalley invariants of $\g$.
\end{thm}
The proof makes use of the so-called $1$-chop map, a Poisson map from $\g$
to $\g'$, which was introduced for $sl(n)$ in Singer [1990, 1993] and
generalized for an arbitrary Lie algebra in Gekhtman and Shapiro [1999].
This map and the functions (\ref{ij})
play a key role in the proof of the complete integrability
of the Toda flows on generic orbits (see Deift {\it et al.} [1986]
and Ercolani {\it et al.} [1993] for the $sl(n)$ case and Gekhtman and
Shapiro [1999] for a general case).

It also follows from Theorem 2.1  that for constructing
Darboux coordinates
on $\O_{m, h_0}$, it is sufficient to construct Darboux coordinates for
a rather simple Poisson algebra generated by (\ref{rel3}).

The following example demonstrates the connection with the peakons lattice.
\medskip

\noindent
{\bf Example 2.2.} Let $\g=sl(n)$. Then $e_m=E_{1n}, e_{-m}=E_{n1},
h_m=E_{11}- E_{nn}$,
$$
\tilde F = \left\{ \zeta=\left [
\begin{array}{ccc} y&0&0\\v_1&0&0\\x&v_2^T&-y\end{array}
\right ]\ :\  x,y\in \R, v_1,v_2 \in \R^{n-2} \right \}\ .
$$
Since Tr$(L)$ is invariant under the action (\ref{coad}), we can
define $S$ to be the space of all symmetric matrices, not necessarily with
zero trace. One can choose $h_0$ to be any diagonal matrix of
the form $h_0=$diag$(\kappa,d,\kappa)$, where $d$ is a  $(n-2)\times(n-2)$
diagonal matrix.
It follows from (\ref{orbit}) that
\begin{equation}
\O_{m, h_0} = \left  \{ L = \left [
\begin{array}{ccc} y+\kappa &v_1^T&x\\v_1&d+ \pi_\S(x^{-1}v_1v_2^T)&v_2\\
x&v_2^T&\kappa-y\end{array}
\right ]\ :\ x>0,  y\in \R, v_1,v_2 \in \R^{n-2}
 \right \}\ .
\label{orbitsl}
\end{equation}

If $d=0$, an open set $\{ -\kappa < y < \kappa; \ v_{1i} >0, \ v_{2i} >0, \ i=1,
\ldots,n-2\}\subset \O_{m, h_0}$ admits Darboux coordinates
$p_i$ and  $q_i$
defined as follows:
$$p_1=y+\kappa, \  p_n=\kappa -y,\
p_{i+1}= \frac{v_{1i}v_{2i}}{x} \ , \ i=1,\ldots, n-2,\ $$
and
$$
\exp(q_n-q_1)=\frac{x^2}{\kappa^2-y^2},\
\exp(q_i-q_1)=\frac{xv_{1i}}{(y+\kappa)v_{2i}}\ , i=2,\ldots, n-1.$$
With this parameterization matrix $L$ in (\ref{orbitsl}) coincides
with the Lax matrix (\ref{laxpl}). Note that $p_1+\cdots +p_n=const$ on
$\O_{m, h_0}$
and that coordinates $q_i$ are defined only up to a translation.
Note that a different choice of Darboux coordinates for $\O_{m,0}$ was proposed
in Kamalin and Perelomov [1985] as an example of  the general construction of
canonical coordinates on coadjoint orbits.

\medskip

\noindent
{\bf Example 2.3.} Let $\g$ be an exceptional algebra of type $G_2$. Denote the
short and long
simple roots of $\g$ by $\nu_1, \nu_2$ resp.,  and root vector
corresponding to a negative root $\alpha = -(i\nu_1 + j\nu_2)$ by
$e_{ij}$.
Then $\F$ is spanned by vectors $ e_{-m}=e_{32}$ and $e_{i1}, \ i=1,\ldots 4$.
Thus, for any $h_0$ the orbit $\O_{m, h_0}$ is $6$-dimensional, while there
are only two independent Chevalley invariants. This illustrates the necessity
of using functions (\ref{ij}) to establish complete integrability.

Put $h_0=0$ and choose Darboux coordinates $q_i,p_i$ for Poisson algebra
(\ref{rel3})
with $N=3$ in the form:
$$q_1=x_0,\ p_1= -\frac{y_0}{x_0},\ q_2=- \frac{x_1}{3\sqrt{x_0}}, \
 p_2=\frac{6 y_1}{\sqrt{x_0}}, \ q_3=- \frac{y_2}{\sqrt{x_0}},\
p_3=- \frac{2 x_2}{\sqrt{x_0}}\ .$$
Then Theorem 2.1 gives the following integrable polynomial Hamiltonian
quadratic in momenta:
\begin{equation}
H(p,q)=\frac{1}{12}(p_3 q_2 + 6 q_3^2)^2 + \frac{1}{8}(p_3 q_3 + p_2 q_2)^2
+ \frac{q_1}{12}(p_3^2 + 12 p_2^2 + 3p_1^2) + q_1^2 + 3 q_3^2 q_1+ 3 q_3^4
\label{g2}
\end{equation}

A detailed analysis of the system generated by (\ref{g2}), as well as
integrable systems associated with peakon-type orbits in other simple Lie
algebras will be given in a forthcoming publication, Alber and Gekhtman
[1999].


\begin{thebibliography}{}

\item Alber, M.S., R. Camassa, Yi. Fedorov, D.D. Holm and J.E. Marsden
[1999], The Geometry
of New Classes of Weak Billiard Solutions of Nonlinear PDE's (subm.), preprint LAUR, Los
Alamos Natl. Lab.

\item Alber, M.S., G.G. Luther, and J.E. Marsden [1997],  Energy Dependent
Schr\H{o}dinger
Operators and Complex Hamiltonian Systems on Riemann Surfaces, {\it
Nonlinearity\/} {\bf
10} 223--242.

\item Alber, M.S., R. Camassa, D.D. Holm and J.E. Marsden [1995], On
the link between umbilic geodesics and soliton solutions of nonlinear PDE's,
{\it Proc. Roy. Soc} {\bf 450} 677-692.

\item Alber, M.S., R. Camassa, D.D. Holm and J.E. Marsden
[1994], The geometry of peaked solitons and billiard
solutions of a class of integrable pde's,
{\it Lett. Math. Phys.\/} {\bf 32}, 137--151.


\item Belokolos, E.D., A.I. Bobenko, V.Z. Enol'sii, A.R. Its, and
V.B. Matveev [1994], {\it Algebro-Geometric Approach to Nonlinear
Integrable Equations.\/} Springer-Verlag series in Nonlinear Dynamics.


\bibitem{ch1} Camassa, R. and D.D. Holm [1993], An integrable
shallow water equation with peaked solitons,
{\it Phys. Rev. Lett.\/}  {\bf 71}, 1661-1664


\item Calogero, F., [1995] An integrable Hamiltonian system,
{\it Phys. Lett. A.} {\bf 201}, 306--310

\item Calogero, F., Francoise, J.-P. [1996], Solvable quantum
version of an integrable Hamiltonian system, {\it J. Math. Phys.}
{\bf 37} (6),  2863--2871


\bibitem{dlnt} P. Deift, L. Li, T. Nanda and  C. Tomei [1986], The Toda flow on
a generic orbit is integrable, {\it Comm. Pure \& Appl. Math.} {\bf 39}
183--232.





\bibitem{efs} N. Ercolani, H.  Flaschka, S. F. Singer [1993], The geometry
of the full Kostant-Toda lattice, {\it Progr.  Math.}  {\bf 115} ,
181--225.





\bibitem{gs2} M. I. Gekhtman and  M. Z. Shapiro [1999], Non-commutative and
commutative
integrability
of generic Toda flows in simple Lie algebra, {\it Comm. Pure $\&$ Appl. Math.}
{\bf 52}, 53-84.



\bibitem{gw}R. Goodman and N. R. Wallach [1984] Classical and quantum
mechanical systems of Toda-lattice type. II. Solutions of the
classical flows, {\it Comm. Math. Phys.} {\bf 94}, 177--217.


\bibitem{hum} J. E. Humphreys [1980],
Introduction to Lie Algebras and Representation
Theory, Springer-Verlag, Berlin.

\bibitem{kamper} S. A. Kamalin and A. M. Perelomov [1985], Construction of
canonical
coordinates on polarized coadjoint orbits of Lie groups, {\it Comm. Math.
Phys.} {\bf 97}, 553-568.



\bibitem{K} B. Kostant [1979],
The solution to a generalized Toda lattice and representation theory,
{\it Adv. Math.} {\bf 34}, 195--338.

\bibitem{ov} Onishchik, A. L. and Vinberg, E. B. [1990],
Lie Groups and Algebraic Groups, Springer-Verlag,
Berlin.



\bibitem{acnjvm} O. Ragnisco and M. Bruschi [1996], Peakons, $r$-matrix and
Toda lattice, {\it Physica A} {\bf 228}, 150--159.



\bibitem{si90} S. Singer [1990],  Doctoral Dissertation.
Courant Institute of Mathematical Sciences. New York University.

\bibitem{si93} S. Singer [1993], Some maps from the full Toda lattice
are Poisson, {\it Phys. Lett. A} {\bf 174} , 66--70.


\bibitem{whitham} Y.B. Suris [1996], A discrete time peakons lattice,
{\it Physics Letters A}  {\bf 217}, 321--329.



\end{thebibliography}
\end{document}